\documentclass[aps,prl,twocolumn,showpacs,superscriptaddress,groupedaddress,nofootinbib]{revtex4}
\usepackage[dvips]{graphicx}
\usepackage{epsfig}
\usepackage{subfigure}
\usepackage{float}
\usepackage{color}
\usepackage{ucs}
\usepackage[utf8x]{inputenc}
\usepackage{soul}
\usepackage{xr-hyper}
\usepackage{amsmath}

\begin{document}
\title{Quadrupole Arrangements and the Ground State of Solid Hydrogen}
\date{\today} \author{Sebastiaan van de Bund} \author{Graeme
  J. Ackland} \affiliation{School of Physics and Centre for Science at
  Extreme Conditions, University of Edinburgh, Edinburgh EH9 3JZ, UK.}


\begin{abstract}
The electric quadrupole-quadrupole ($\mathcal{E}_{qq}$) interaction is believed to play an important role in the broken symmetry transition from Phase I to II in solid hydrogen. To evaluate this, we study structures adopted by purely classical quadrupoles using Markov Chain Monte Carlo simulations of fcc and hcp quadrupolar lattices.  Both undergo first-order phase transitions from rotationally ordered to disordered structures, as indicated by a discontinuity in both quadrupole interaction energy ($\mathcal{E}_{qq}$) and its heat capacity.  Cooling fcc reliably induced a transition to the P$a3$ structure, whereas cooling hcp gave inconsistent, frustrated and $c/a$-ratio-dependent broken symmetry states. Analysing the lowest-energy hcp states using simulated annealing, we found P$6_3/m$ and P$ca2_1$ structures found previously as minimum-energy structures in full electronic structure calculations.  The candidate structures for hydrogen Phases III-V were not observed. This demonstrates that $\mathcal{E}_{qq}$ is the dominant interaction  determining the symmetry breaking in Phase II.  The disorder transition occurs at significantly lower temperature in hcp than fcc, showing that the $\mathcal{E}_{qq}$ cannot be responsible for hydrogen Phase II being based on hcp.\\
\end{abstract}

\maketitle

\section{Introduction}
Hydrogen is the simplest element, yet even in its solid molecular form it exhibits surprising and complicated behavior. Solid phases occur via induced dipole-dipole (van der Waals) and quadrupolar attractions. The first of these is Phase I, a molecular solid first created in the laboratory in 1899 by James Dewar\cite{dewar-hydrogen}. It is now known to be a quantum molecular solid, where the hydrogen molecules themselves are free to rotate, with their centres of mass on hexagonal close packed lattice sites\cite{loubeyre-phase1-xray}.  At higher pressure and low temperature, the rotation ceases and the system enters the ``broken symmetry'' Phase II\cite{hemley1988phase,lorenzana1989evidence,lorenzana1990orientational}. 
Although the exact structure of hydrogen Phase II is not known, it does have an optical Raman mode which seems to correspond to the layer-mode in Phase I. This, and X-ray evidence, suggests an hcp-based structure, but the orientation of the molecules is unknown.

Solid nitrogen has a similar hcp-based molecular solid with free
rotors interacting via a quadrupole moment.
At ambient pressure, it exists below 63K\cite{schuch1970crystal}, transforming to a broken symmetry phase with cubic P$a3$ symmetry below 36K.

A number of structures have been posited for hydrogen Phase II, all of which can be regarded as molecules on an hcp lattice and differing in the molecular orientation. In 1991 using density functional theory, Kaxiras and coworkers predicted metallization in a 2-atom cell\cite{kaxiras1991onset}, then showed that a 4-atom P$mc2_1$ supercell is more stable and non-metallic\cite{kaxiras1992energetics}. In 1992 Nagara and Nakamura \cite{nagara1992stable}, proposed an 8-atom P2$_1$/c, then in 1997 Kohanoff\cite{kohanoff1997solid} found a more stable 8-atom P$ca2_1$, a record twice beaten by Pickard and Needs (16-atom P$6_3/m$\cite{pickard2007structure} and 24-atom P$2_1/c$ \cite{pickard2009structures}).  The relative stability of these structures is sensitive to details of the calculation, and the trend towards greater stability with ever-larger supercell implies that no consensus exists on the most stable phase.

Under further pressurization, a strong Infrared signal heralded another non-centrosymetric broken symmetry Phase III\cite{hemley1988phase,lorenzana1989evidence}.  Several other molecular solid phases have also been observed\cite{eremets2011conductive,howie2012mixed,zha2013high,dalladay2016evidence}, yet apart from Phase I, none of the crystal structures are unambiguously known. Numerous possibilities are advanced by density functional theory\cite{pickard2007structure,pickard-phaseiv,magdau2013identification,magdau2017simple,pickard2012density,liu2012room,geng2012high}, while X-ray measurements have only recently started to probe such high pressures, indicating an underlying hcp structure for phases I-IV\cite{ji2019ultrahigh}.
At still higher pressures, the situation is controversial, with claims of conductive molecular and metallic atomic phases, but these are beyond the scope of the current work.

By contrast, the case of low temperature solid N$_2$ is much simpler: it adopts the P$a3$ structure based on an fcc lattice. A different fcc-based oC4 structure was reported for Cl$_2$, similar to Br$_2$ and I$_2$\cite{dalladay2019band,fujihisa1995structural}.

The DFT-predicted structures are sensitive to details of the approximations and treatment of various physical effects: nuclear quantum effects, electron exchange and correlation, van der Waals interactions etc, and it now seems unlikely that an unambiguous answer will be found.  Here we take an alternate approach which resolves a fundamental issue in physics, and illuminates the understanding of hydrogen Phases I and II: we examine a single effect, the quadrupole-quadrupole interaction, and ask what structures it would prefer.

Lattice Monte Carlo is a standard technique for tackling phase transitions. The lattice removes the degrees of freedom associated with positions and momenta $\vec{r}_i$ and $\vec{p}_i$ of the molecules, and enables us to focus on just one effect, molecular orientation on an underlying fcc or hcp lattice.  The aim of this work is twofold, and applies two types of Monte Carlo run. First, simulated annealing is used to identify the ground state. Second, finite temperature runs are used for gathering statistics and identifying the broken-symmetry transition.

\section{Monte Carlo Methods}

We consider a linear quadrupole oriented in 3D  on each lattice site $\vec{\sigma}(\theta, \phi)$. Each quadrupole interacts with its neighbors according to standard electrostatics \cite[p.51]{stone-imf}
with angular dependence

\begin{equation} \label{eq:eqq-orient}
\begin{split}
\Gamma(\vec{\sigma}_i, \vec{\sigma}_j, \hat{\vec{R}}) &= 35 (\vec{\sigma}_i\cdot\hat{\vec{R}})^2(\vec{\sigma}_j\cdot\hat{\vec{R}})^2 -5(\vec{\sigma}_i\cdot\hat{\vec{R}})^2
-5(\vec{\sigma}_j\cdot\hat{\vec{R}})^2\\
&\,\,\,+2(\vec{\sigma}_i\cdot\vec{\sigma}_j)^2
-20(\vec{\sigma}_i\cdot\hat{\vec{R}})(\vec{\sigma}_j\cdot\hat{\vec{R}}) (\vec{\sigma}_i\cdot\vec{\sigma}_j) + 1
\end{split}
\end{equation}

\begin{figure}
\centering
\includegraphics[scale=1.8]{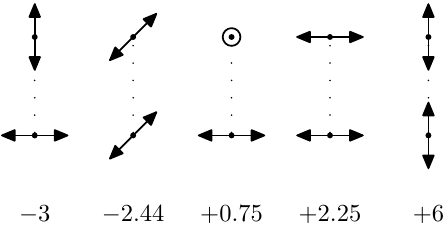}
\caption{A selection of favorable and unfavorable orientations of two point quadrupoles (with positive quadrupole moment $Q$) adapted from ref. \cite[p.53]{stone-imf}. Numbers correspond to the orientational factor $\Gamma$.}
\label{fig:eqq-orientational}
\end{figure}

Examples of this somewhat cumbersome expression are shown in 
Fig. \ref{fig:eqq-orientational}. 
The quadrupole-quadrupole energy, denoted by $\mathcal{E}_{qq}$, depends on the orientations $\vec{\sigma}$ as well as the separation vector between molecules. This complication precludes straightforward analysis.

The total quadrupole-quadrupole energy can be written as

\begin{equation}
 \mathcal E_{qq} = \sum_{ij} \mathcal{E}_{ij} = \frac{J}{p_{ij}^5}\Gamma(\vec{\sigma}_i, \vec{\sigma}_j, \hat{\vec{R}}_{ij}), \qquad
J \equiv \frac{3Q^2}{4\pi\epsilon_0 R^5}
\end{equation} 

where $R$ is a fixed length scale (usually the nearest neighbor distance in the lattice), such that the distances between a pair of quadrupoles can be written as $r_{ij} = p_{ij}R$. The energy scale $J$ dedimensionalizes the simulation, depends on which element is considered, and should not be confused with Joules. Energies will generally be quoted in units of $J$, temperatures in units of $J/k_B$, and heat capacities in units of $J/k_B^2$. For clarity, values of dimensionless temperatures will be denoted by $T^*$. This then allows for rescaling values to any desired volume by evaluating the constant $J$ using the appropriate value of $R$ to match the required volume.  For H$_2$, the quadrupole moment is taken as $Q = 0.26$D{\AA}\cite{simple-hydrogen,orcutt-qmoment} where $D$ is the Debye unit.

The total energy of the system includes many other effects, including covalent bonding, van der Waals interactions, etc. These are larger than $\mathcal{E}_{qq}$, but we will assume that they act between molecules, independent of the molecular orientation.  Thus we presume that the hcp structure of Phases I and II of hydrogen arises from central forces, and the only orientational-dependent interaction is from $\mathcal{E}_{qq}$. 

Ensemble averages were calculated in a canonical lattice Monte Carlo simulation using importance sampling via the Metropolis-Hastings algorithm\cite{hastings-mcmc,metropolis-equation} based on single-site rotations within a cone of tunable size.  We devised an algorithm to dynamically tune the size of the random update. 

The maximum allowed rotation angle was adjusted during the equilibration to achieve a Metropolis acceptance ratio of $0.5$, then fixed for the longer measurement run.

While effective, this clearly impedes the ability to overcome energetic barriers at lower temperatures. An additional algorithm was devised for simulated annealing of hcp specifically. The single rotor Metropolis is exceedingly slow at reconfiguring layered structures, which is the main feature at high $c/a$ ratios. Hence, in order to obtain the hcp ground state, an additional move is proposed where an entire layer is moved in the $\mathbf{a}-\mathbf{b}$-plane along a random choice of the six shortest in-plane translation vectors, with the usual Metropolis-Hastings acceptance condition. When performing simulated annealing, the temperature for this Metropolis move does not need to be equal to that used for the single rotors, so it is tuned to optimize acceptance ratios.

\subsection{Correlation Functions}

To determine appropriate measurement intervals, we monitored the quadrupole-quadrupole energy correlation function

\begin{equation} \label{eq:autocorrelation}
C_{qq}(t) = \frac{\langle \mathcal{E}_{qq}(t^\prime+t)\mathcal{E}_{qq}(t^\prime) \rangle - \langle  \mathcal{E}_{qq}(t^\prime)\rangle^2}{\langle \mathcal{E}_{qq}(t^\prime)^2\rangle-\langle \mathcal{E}_{qq}(t^\prime)\rangle^2}
\end{equation}

where the averages are over MC sweeps $t^\prime$. This becomes zero if the $\mathcal{E}_{qq}$ were fully uncorrelated, and 1 if fully correlated.

The integrated autocorrelation time is then given by
\begin{equation}
    \tau_\text{int} = \frac{1}{2} + \sum_{t=1}^M C_{qq}(t)
\end{equation}
$C_{qq}(t)$ often decays exponentially, but is dominated by noise at high $t$ which means the sum becomes unreliable. Hence, $\tau_\text{int}$ was estimated by limiting the sum to a window. $M$ was tuned to be the lowest value for which the condition $M\leq 6\tau_\text{int}(M)$ holds true.

\subsection{Thermal Averaging}

The $\mathcal{E}_{qq}$ contribution to heat capacity per rotor at constant volume are calculated from the fluctuations in the sampled energies\cite{welford-var}\cite[p.141]{tuckerman-statistical}:

\begin{equation}
c_V = \frac{1}{T^2}(\langle  \mathcal{E}_{qq}^2\rangle - \langle  \mathcal{E}_{qq}\rangle^2)
\end{equation}

The sampling error is estimated using\cite[p.93]{landau-guide}

\begin{equation}
\sigma_\text{eqq} = \sqrt{\frac{\langle \mathcal{E}_{qq}^2 \rangle -\langle \mathcal{E}_{qq} \rangle^2}{n\tau} 2\tau_\text{int}}
\end{equation}

where $\tau_\text{int}$ is the integrated autocorrelation time and the interval between measurements is $\tau$ sweeps. The term $\frac{2\tau_\text{int}}{n\tau}$ can be thought of as the number of effective, statistically independent measurements.

\par The uncertainty in the heat capacity is found through resampling\cite{johnson-bootstrap}: we recalculate the heat capacity 1000 times, using $n$ measurements sampled from the list of $\mathcal{E}_{qq}$ energies, with repetition being allowed.  This gives 1000 estimates of heat capacity $\tilde{\sigma}$, and our uncertainty is the standard deviation of these recalculated values.

\subsection{Choice of Lattice}
Two types of lattice were considered in the Monte Carlo runs: hexagonal close packed lattice, as observed in hydrogen Phase I, and the close packed fcc lattice. In both cases there is frustration: it is not possible to minimize $\mathcal{E}_{ij}$ for every pair of molecules simultaneously.  A structure with P$a3$ symmetry is believed to minimize the $\mathcal{E}_{qq}$ energy on an fcc lattice\cite{miyagi-eqqfcc}, but the minimum energy for hcp is not known. For ideal $c/a$ ratio, hcp and fcc have the same packing fraction and the same number of nearest and next-nearest neighbors, being 12 and 6 respectively.

Because the quadrupole coupling drops off as $\sim r^{-5}$, the interaction can be truncated.  In fcc, truncation after second neighbors already stabilizes P$a3$ and further interactions only cause small shifts to the transition temperature. However, for hcp  it was found that beyond next-nearest neighbor interactions were important and to converge the calculations around 140 neighbors (out to four lattice constants $a$) were required. The effect of changing the $c/a$ ratio in hcp at constant density was also investigated.

The choice of lattice constant, and therefore volume, only affects $\mathcal{E}_{qq}$ up to a constant since $\mathcal{E}_{qq}\sim r^{-5}$. All length scales can therefore be dedimensionalized using the length scale $R$. This can be reintroduced later to obtain $\mathcal{E}_{qq}$ at the desired volume.

\section{Results \& Discussion}
\subsection{Ground States}
\subsubsection{Triangular Lattice}

\par
\begin{figure}[h!]
\centering
\includegraphics[scale=0.5]{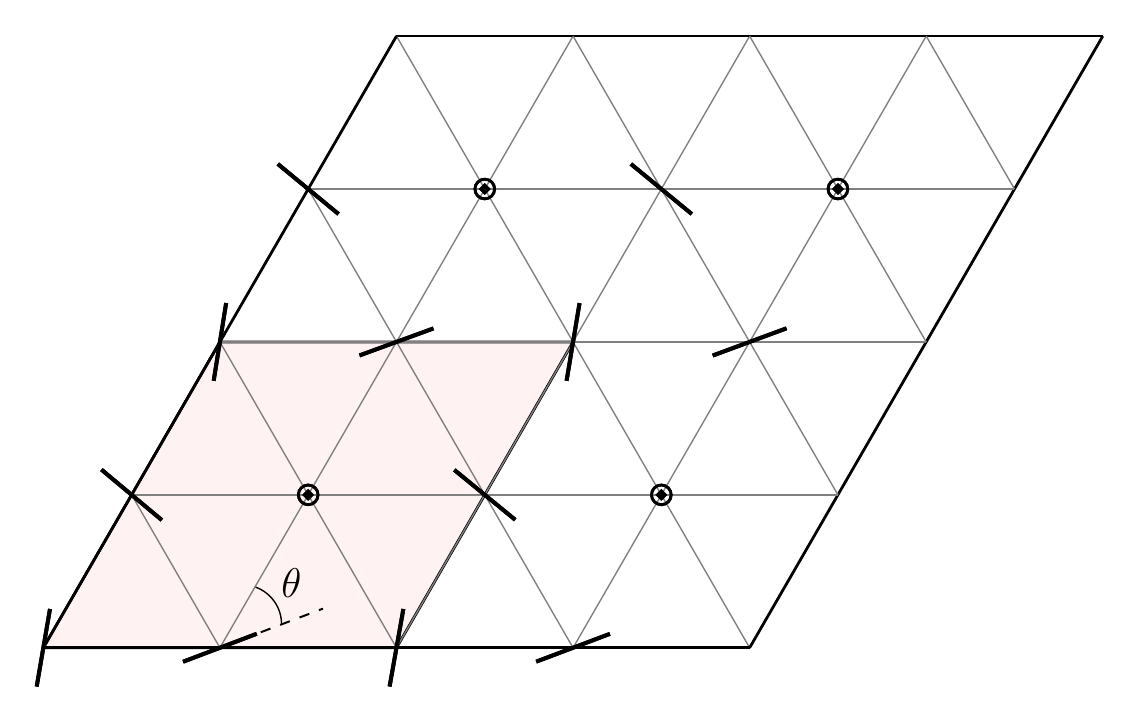}
\caption{Ideal unit cell arrangement of quadrupole rotors on a triangular lattice. The main motif consists of a central vertical rotor surrounded by six planar rotors with no frustration. This structure has P6 symmetry.}
\label{fig:triangular}
\end{figure}

 Both fcc and hcp consist of layers of triangular lattices, differing only in the stacking order, so we start by determining the stable configuration of a single layer using the MC procedure quenched to low $T$. The stable unit cell consists of 4 rotors (Fig. \ref{fig:triangular}), one perpendicular to the plane and three rotors parallel to the plane. If the unit cell is repeated, a 6-fold pattern about the vertical rotor is found, where the most favorable orientation of the plane-parallel rotors is $\theta=\pm$ 0.669 rad with respect to the vector connecting itself with the central plane-perpendicular rotor. The spacegroup of this layer structure is P6.
Attempting to stack these layers in hcp ABA fashion leads to significant frustration, as will be discussed.

\subsubsection{fcc Lattice}

\begin{figure}
\centering
\includegraphics[scale=1.4]{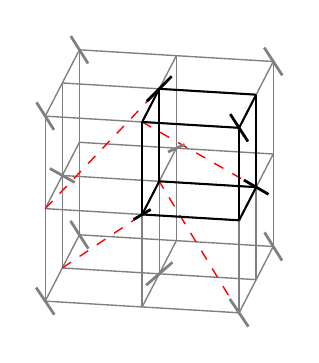}
\hspace{2em}
\caption{$T^* \sim 0$ structure upon cooling of disordered fcc
in the conventional unit cell representation. Red dashed lines are a guide to the eye. The rotors do not lie in the close-packed plane like in the ground state triangular lattice.}
\label{fig:fcc-Pa3}
\end{figure}

We locate the fcc stable ground state using simulated annealing by linearly lowering the temperature from $T^*=10$ to $T^*=0.0001$ over the course of 5$\times 10^5$ sweeps. All supercell sizes larger than 4x4x4 (256 molecules) consistently give the same structure, P$a3$, as  shown in Fig. \ref{fig:fcc-Pa3}.

This P$a3$ symmetry structure can be thought of as four interpenetrating cubic lattices with the rotors pointing towards the far corner of a neighboring cubic cell\cite[p.682]{mao-ultrahigh}. It has been widely claimed to minimize the lattice $\mathcal{E}_{qq}$ energy\cite{miyagi-eqqfcc}, though careful reading of that work proves only that it is a local minimum - because of frustration it is possible obtain lower energy for an individual molecule, but only at the expense of increased energy elsewhere. The absence of any competing minima in the simulation here provides further support for the assumption that it is a global minimum.

\subsubsection{hcp Lattice}

Compared with fcc, hcp has an additional complication due to its $c/a$ ratio. In real molecular solids, the $c/a$ ratio is not the close-packing value $\sqrt{8/3}$. Consequently, a range of $c/a$ ratios was considered in order to identify the ground state of the rotors on an hcp lattice. The unit cell volumes were kept constant and equal to 1 throughout to allow for rescaling to relevant units at a later stage.

Changing the $c/a$ ratio also changes the nearest neighbor distance. In order to keep the energy scales comparable for different $c/a$ values, the coupling constant $J$ uses the nearest neighbor distance in the ideal $c/a$ ratio of $\sqrt{8/3}$ at unit volume as a common length scale throughout. This is equal to $R\approx 0.891$ in the corresponding unit of distance.

A range of $c/a$ values between 1.2 and 2.2 with an interval of 0.004 was considered. Simulated annealing was performed by linearly cooling from $T^*=5$ to $T^*=0$ over the course of 2$\times 10^4$ sweeps. This was done for four supercells of sizes 2x2x1, 4x4x2, 4x4x3 and 6x6x3. With two sites per primitive unit cell, these system sizes range from 8 to 216 rotors. Because of the large number of metastable minima, 10 separate simulations per $c/a$ value were performed.

It was found that many neighbors beyond nearest and next-nearest were necessary to distinguish competing structures. To ensure convergence, the cutoff was set equivalent to 4 lattice spacings $a$ of the ideal $c/a$, which included over 140 neighbors for each quadrupole.

\begin{figure}[h!]
\centering
\includegraphics[scale=0.75]{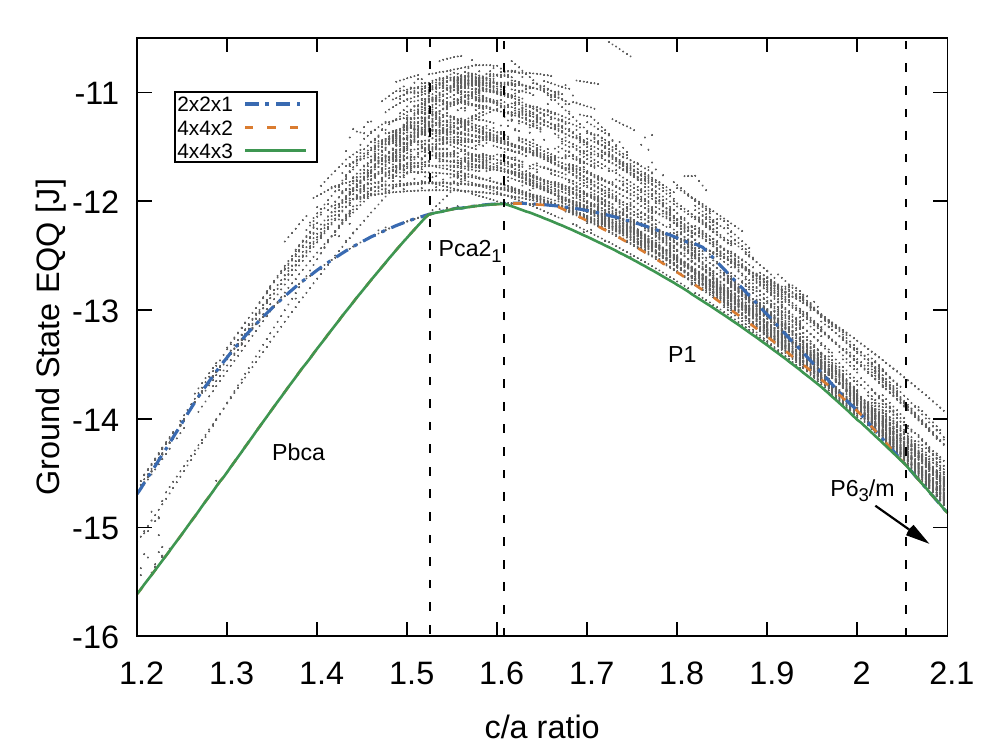}
\caption{$T^*\sim 0$ energies per rotor obtained by simulated annealing of disordered hcp for all three supercells. The lines indicate the lowest obtained energies for each supercell, from which intersections distinguish the various phases. The 6x6x3 data did not produce any lower energy structures.}
\label{fig:hcpgsenergy}
\end{figure}

Finally, in order to obtain clearer ``phase lines'', each final result of simulated annealing was used as a starting configuration for runs at other $c/a$ ratios. This procedure gives multiple continuous lines in Fig. \ref{fig:hcpgsenergy} where the simulated annealing did not obtain the lowest energy structure.

The obtained ground state energies (per rotor) are shown in Fig. \ref{fig:hcpgsenergy}. The frustration inherent in hcp is evidenced by the many closely spaced lines at higher energies than the main lowest energy line. This can be understood from the fact that the energy landscape from the orientational factor Eq. \ref{eq:eqq-orient} has many local minima, and is difficult to minimize for the whole lattice.

\subsubsection{Low and Near-Ideal $c/a$}

At $c/a$ far below the ideal value, the lowest energy structure has P$bca$ symmetry with 8 rotors in the unit cell. In this regime, the system forms long chains in which the rotors are approximately at an angle of $\pi/2$ with respect to their separation vector. This 2x4x1 unit cell was found to have the lowest energy up until $c/a\sim1.5$.

\begin{figure}[h!]
\centering
\begin{tabular}{cc}
\includegraphics[scale=0.42]{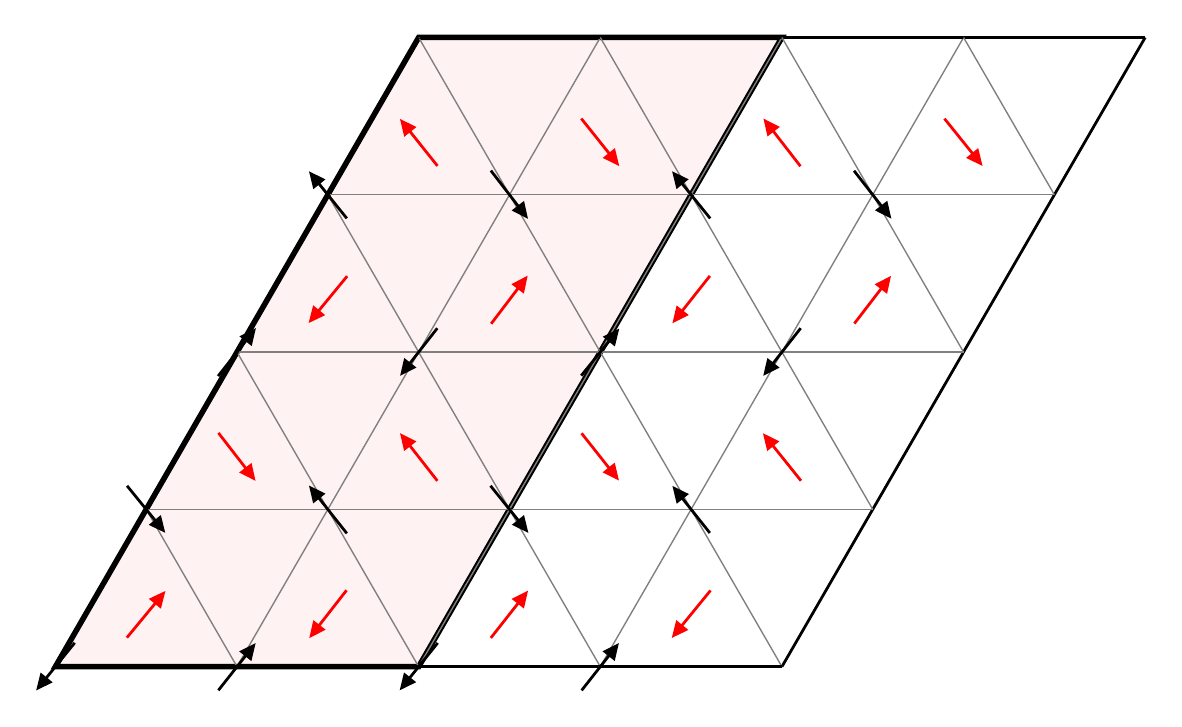} &
\hspace*{-5em}
\includegraphics[scale=0.42]{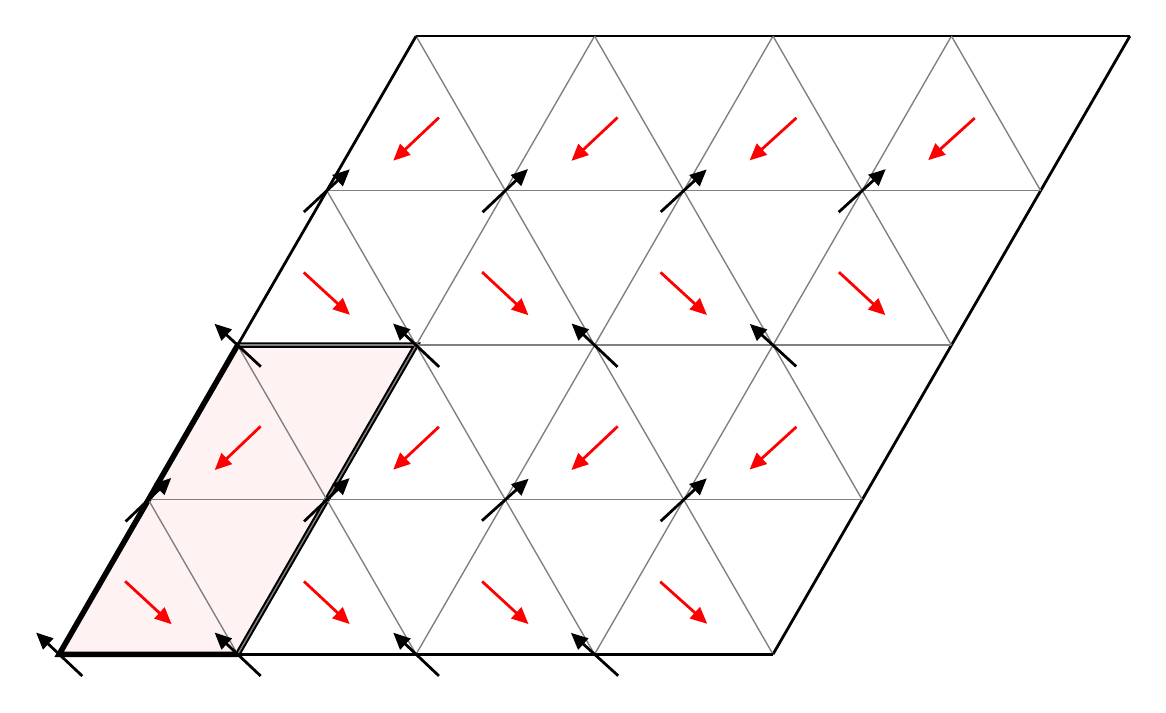} \\
\end{tabular}
\caption{Ground states for low to intermediate $c/a$ ratios. Left is at $c/a$=1.47 and spacegroup P$bca$, while right is at $c/a$=1.55 and spacegroup P$ca$2$_1$. Note that these structures are not planar: arrows indicate the direction of the rotor that points out of the page. Red arrows correspond to the next layer stacked in the $\mathbf{c}$ direction.}
\label{fig:low-ca}
\end{figure}

As the $c/a$ ratio is increased, beyond $c/a\sim 1.5$ a new stable structure with spacegroup P$ca2_1$ appears with a smaller 1x2x1 unit cell. This is also shown in Fig. \ref{fig:low-ca}. The ordering is somewhat similar to P$bca$ in that it favors aligning at $\pi/4$ with respect to the separation vectors. Finally, the curves for  P$ca2_1$ and P$bca$ in Fig. \ref{fig:hcpgsenergy} can be identified as extending beyond their crossover, showing that these structures remain metastable.

\subsubsection{High $c/a$}

Only at very high $c/a$ values beyond around $c/a=2.05$ are the stacked layers sufficiently decoupled such that the ground state is a simple ABA hcp stacking of the stable 2D triangular structure (Fig. \ref{fig:triangular}). This type of stacking is very similar to that found in one of the DFT candidates for Phase II of hydrogen\cite{pickard2007structure,pickard2009structures}, P6$_3/m$.

\begin{figure}[h!]
\centering
\begin{tabular}{cc}
\includegraphics[scale=0.45]{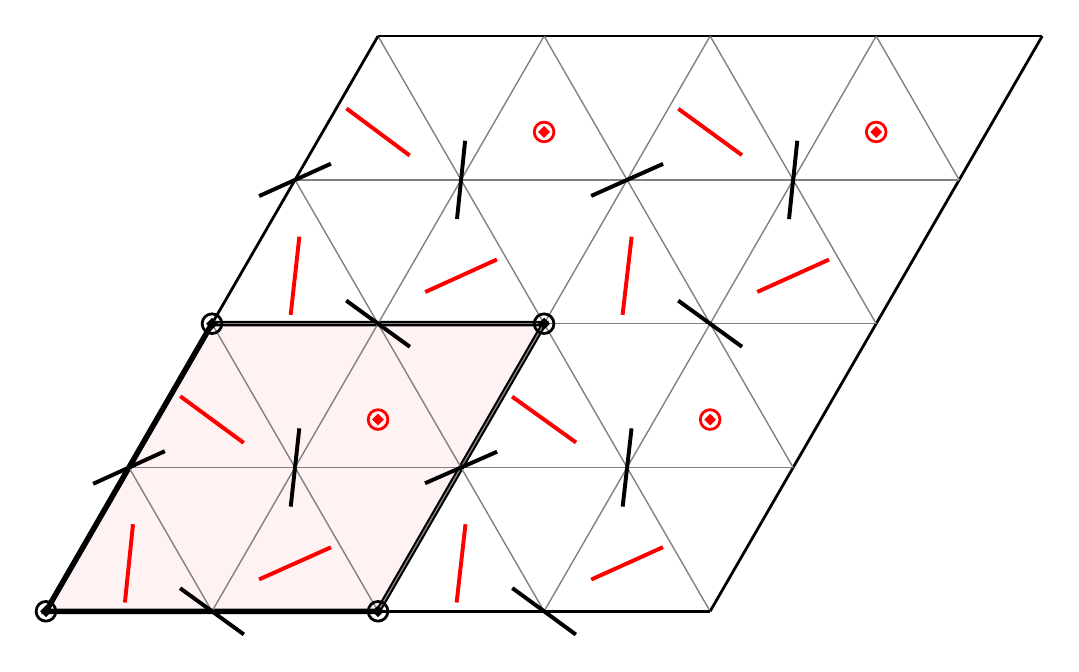} &
\hspace*{-5em}
\includegraphics[scale=0.45]{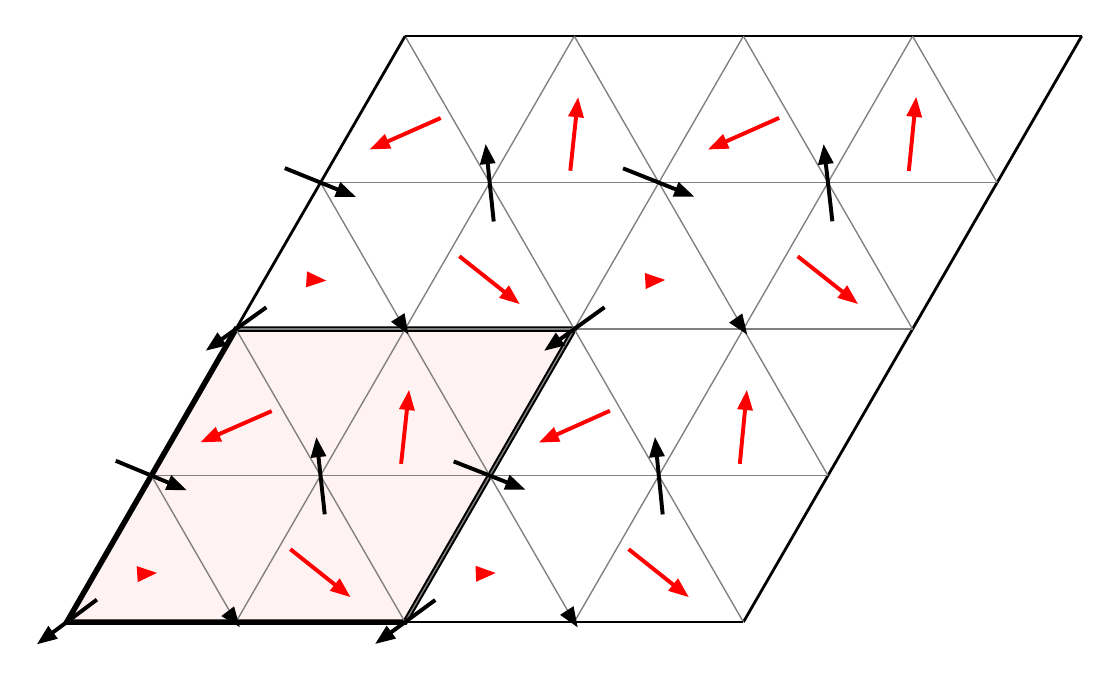}
\end{tabular}
\caption{Two possible stackings of triangular layers observed at $c/a> 1.61$. Left is stacking in P6$_3/m$, which is only stable at very high $c/a$. Right is another possible stacking where the distorted layers still resemble those of P6$_3/m$. While the structure as a whole loses all symmetry, the individual layers possess P-1 symmetry. Circles represent vertical rotors, while single triangles are near-vertical. Red arrows correspond to the next layer stacked in the $\mathbf{c}$ direction.}
\label{fig:high-ca}
\end{figure}

The remaining region between $c/a=1.61$ and $2.05$ is more difficult to characterize. The main motif in these structures is still similar, but with two main differences. Firstly, the stacking is generally no longer ABA. There are effectively four choices for stacking successive P6 layers since the unit cell of the P6 layer is 2x2. An example of two such stackings is given in Fig. \ref{fig:high-ca}.

The second difference is that the P6 layers distort such that the central rotor tilts away from the $\vec{c}$ axis and the planar rotors tilt out of the $\mathbf{a}-\mathbf{b}$ plane. This can in part be explained by the vertical rotors two layers apart pointing towards each other in P6$_3/m$, which is the most unfavorable arrangement(Fig. \ref{fig:eqq-orientational}). Stacking these P6-like layers effectively removes all symmetry of the structure as a whole, so will be labelled as P1.

This problem becomes worse as the number of stacked layers increases. Assuming that the vertical rotor will occupy one of the four sites of each 2x2 layer unit cell, there are already $4^7\sim 16,000$ possible stacking combinations in the 2x2x4 supercell (accounting for translational invariance). Even when allowing for an additional Metropolis-Hastings update where an entire layer is translated in the $\mathbf{a}-\mathbf{b}$ plane, obtaining an unambiguous ground state remains difficult as many of the stackings are similar in energy and the exact distortion of the layers likely depends on the specific stacking order.

\begin{figure}[h!]
    \centering
    \includegraphics[scale=0.425]{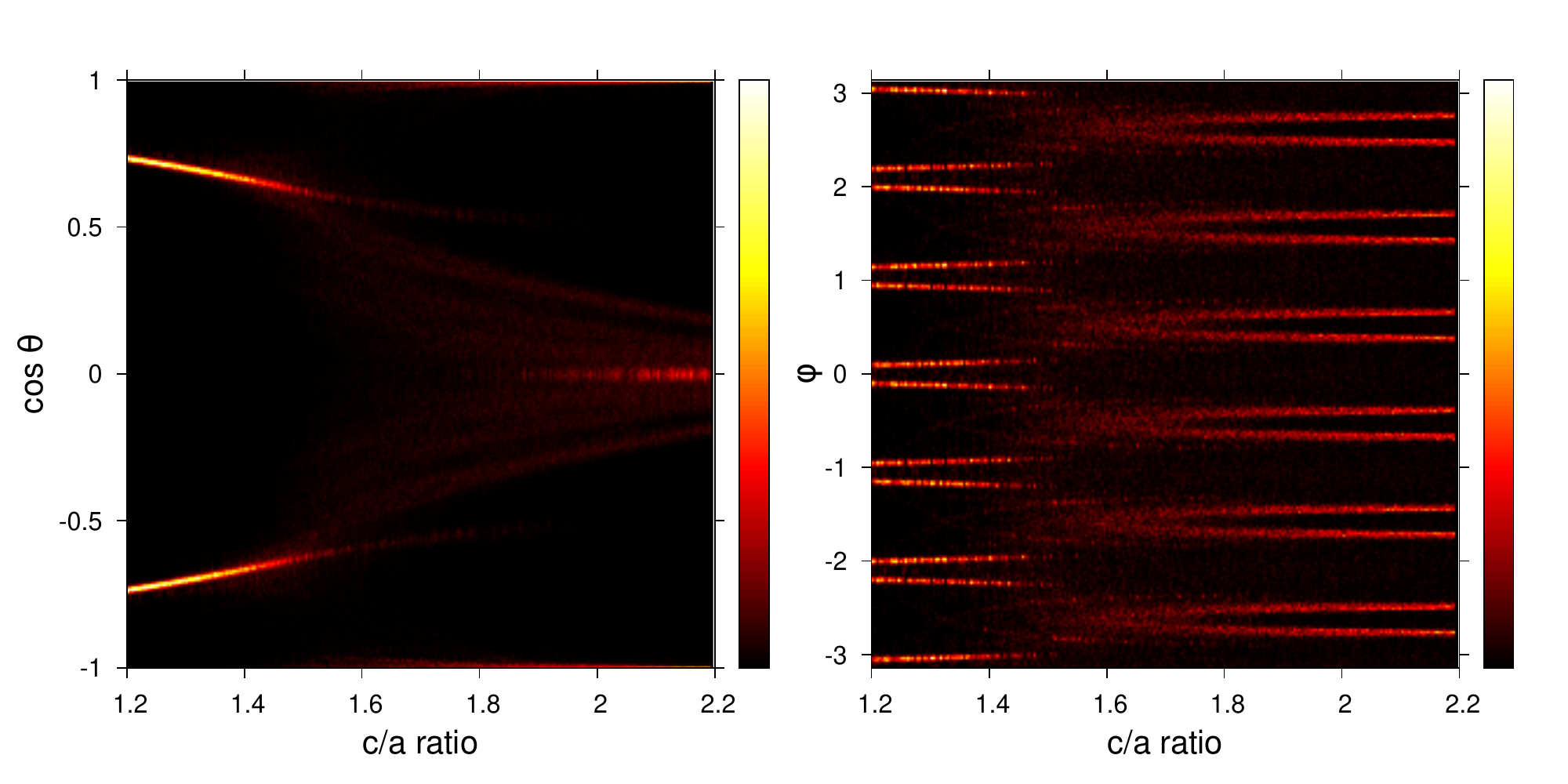}
    \caption{Histograms of $\cos\theta$ and $\phi$ for all supercells, where $\theta$ and $\phi$ are polar and azimuthal angles, as a function $c/a$ ratio.}
    \label{fig:histogram}
\end{figure}


Fig. \ref{fig:histogram} shows the distribution of polar and azimuthal angles of all structures found in Fig. \ref{fig:hcpgsenergy} as a function of increasing $c/a$ ratio. $\theta$ can be seen to start out of the plane, and tend towards largely planar $\pi/2$ orientations at high $c/a$ ratios. Close to ideal c/a, a wide range of different angles are found, reflecting the orientational frustration.

In the intermediate region where the P1 structure is most stable, the angles $\phi$ adopt six-fold rotational symmetry, which supports the notion that the layers are just distorted versions of the planar P6 structure. Because the rotors are tilted out of the plane however, the individual layers generally only possess P-1 symmetry as can be seen in Fig. \ref{fig:high-ca}. The non-ABA stacking removes further symmetries of the structure as a whole.

Finally, since the planar structure in Fig. \ref{fig:triangular} is enantiomorphic, there are two possible sets of the six planar angles $\phi$ shifted relatively by 0.669 rad, which is indeed recovered in the high $c/a$ limit where P6$_3/m$ becomes stable.

\subsubsection{Ground State Energies}

\begin{table}[h!]
\begin{center}
 \begin{tabular}{||c | c c c c c||} 
 \hline
 & P$bca$ & P$ca2_1$ & P1 & P6$_3/m$ & P$a3$ \\ [0.5ex] 
 \hline\hline
 $c/a$ ratio & 1.4 & 1.6 & 1.75 & 2.1 & --\\ [1ex]
 \hline
 $\mathcal{E}_{qq}$/rotor  (meV) & -127.6 & -114.9 & -119.9 & -142.0 & 137.8\\ [1ex]
 \hline
\end{tabular}
\end{center}
\caption{$\mathcal{E}_{qq}$ per rotor for the various phases, calculated using the volume from a Vinet EOS\cite{loubeyre-phase1-xray} at 100GPa.}
\label{tab:gs-eqq}
\end{table}

$\mathcal{E}_{qq}$ can be evaluated for the typical unit cell volumes expected for Phase I and II. Using a Vinet EOS\cite{loubeyre-phase1-xray}, the equilibrium volume for Phase I can be calculated at 100 GPa. These are shown in Table \ref{tab:gs-eqq} for a selection of $c/a$ ratios. Even though some of the obtained structures are stable at a wide range of $c/a$ values (as can be seen from Fig. \ref{fig:hcpgsenergy}), the overall energy scale still remains on the order of 0.1 eV. P6$_3/m$ can be seen to have the lowest $\mathcal{E}_{qq}$, but this is only because of its very high $c/a$ ratio where the planar quadrupoles are much closer to each other.

\subsection{Phase Transitions}

Our model exhibits an order-disorder transition, akin to the Phase I - II ``broken symmetry'' transition in hydrogen, where the rotors go from free rotation at high temperature to strongly inhibited libration with a fixed average orientation at low temperature. We study this by heating from the previously determined ground states. To detect the transition we monitor the ``heat capacity'' contribution as calculated from fluctuations in the $\mathcal{E}_{qq}$ energy. This should diverge at a first-order transition.




Both fcc and hcp ground state structures were heated from their respective ground structure. For fcc, this was the stable P$a3$. For hcp we consider the four ground state structures discussed previously: these are P$bca$, P$ca2_1$, the low-symmetry P1 and finally P6$_3/m$. Because calculating $\mathcal{E}_{qq}$ is relatively expensive, nearest and next-nearest neighbors are considered for all runs which is sufficient to give a clear phase transition.

\begin{figure}[h!]
\centering
\begin{tabular}{cc}
\includegraphics[width=0.51\columnwidth]{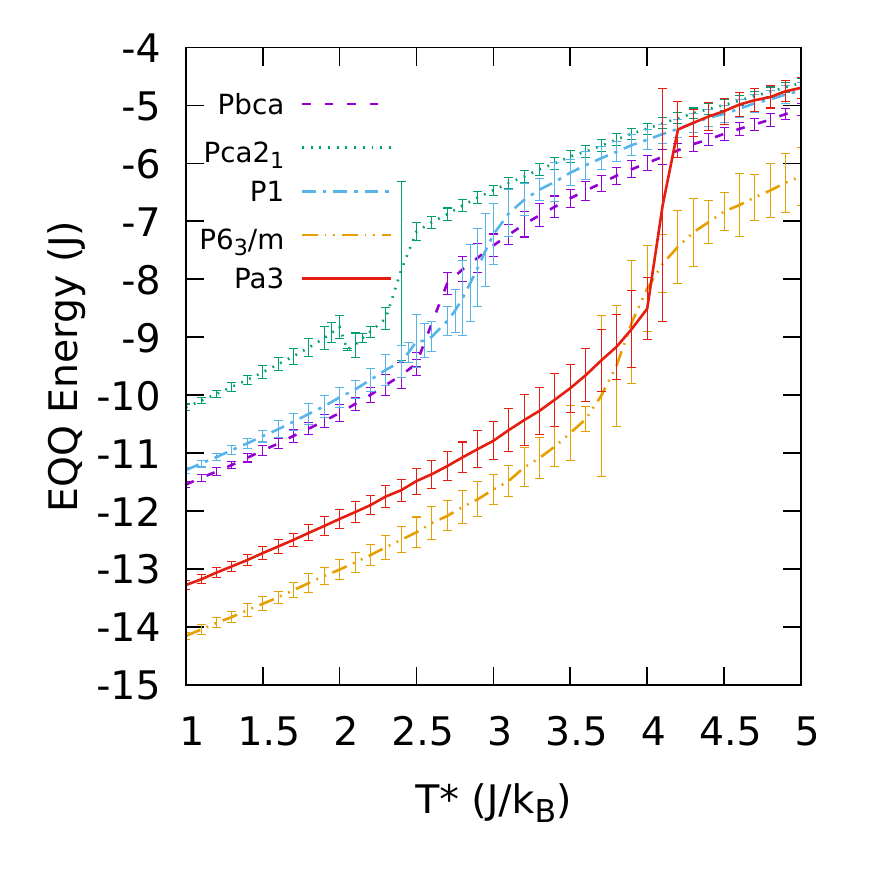} 
& \hspace{-1.5em}
\includegraphics[width=0.51\columnwidth]{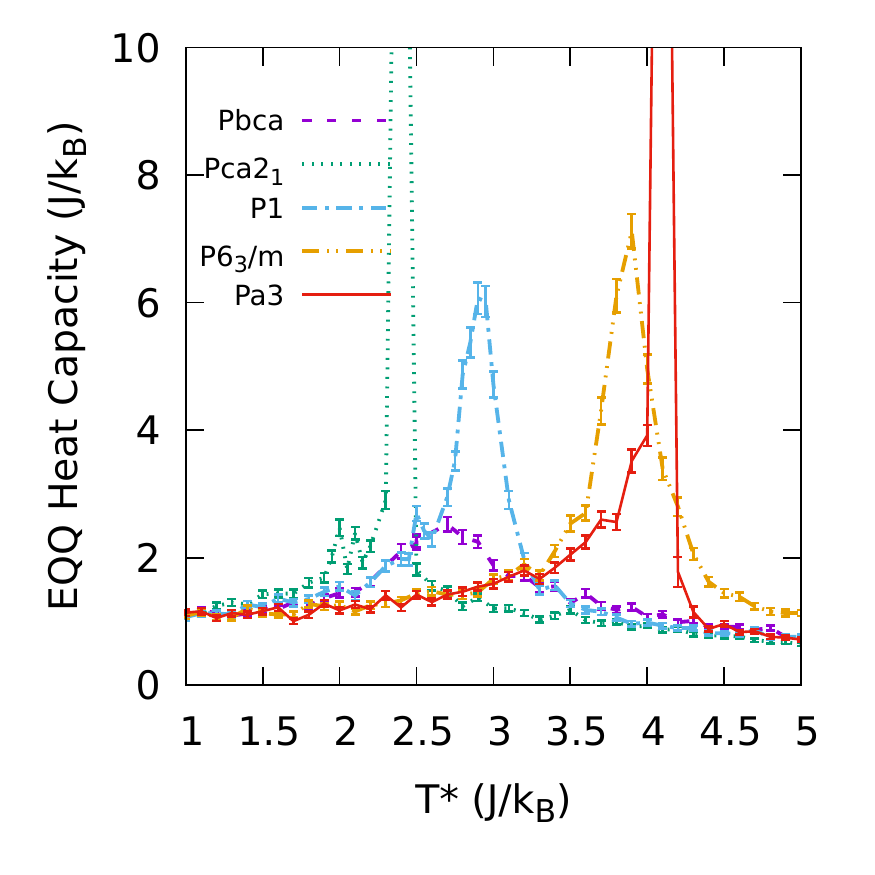} \\
\end{tabular}
\caption{Heating of the P$a3$ fcc ground state and the four hcp ground state structures. Runs for P$bca$, P$ca2_1$, P1 and P$6_3/m$ were performed at $c/a$ ratios of 1.4, 1.6, 1.75 and 2.1 respectively. All are observed to undergo a first-order phase transition. While the critical temperature varies among the structures, the low- and high-temperature heat capacities are similar. Because the autocorrelation time diverges around the critical point, the error bars in that region are likely underestimated.}
\label{fig:hcp-energetics}
\end{figure}

The main tuning lies in the measurement interval $\tau_\text{meas}$ (in MC sweeps), which should ideally be greater than the autocorrelation time. All temperatures were equilibrated and simulated separately. Equilibration and measurement were fixed to 100$\tau_\text{meas}$ and 1000$\tau_\text{meas}$ respectively, while $\tau_\text{meas}$ was tuned to give reasonable error bars. The results for the five structures are shown in Fig. \ref{fig:hcp-energetics}.

The slope of $\mathcal{E}_{qq}$ and therefore the heat capacity in the ordered regime is roughly the same, which is to be expected given their common interaction mechanism. Also, phase transitions below the order-disorder critical temperature can occur. This is seen in the curve for P$ca2_1$ where it transitions to a structure of lower $\mathcal{E}_{qq}$.

P6$_3/m$ proved to be metastable near and below the ideal $c/a$ ratio, and transformed to a less ordered type of structure of the type discussed earlier. Detailed observation of Monte Carlo runs show that the process is gradual, with progressive breaking of the stacking as the temperature is increased. Since each transition lowers the energy, we conclude that P6$_3/m$ is not the ground state at this $c/a$ ratio, but we were unable to identify a unique stable structure.

The hcp disorder transformation temperature is generally significantly lower than in fcc, except for in P6$_3$/m. This does not mean that fcc is more stable, because the lattice stability is determined by angle-independent contributions not included here. Nevertheless, in the context of the current model this can be understood from their relatively low $\mathcal{E}_{qq}$.



The single rotor Metropolis update becomes inefficient near the critical temperature. Sampling this critical region is expensive due to the complicated expression for $\mathcal{E}_{qq}$, so larger systems were not considered. Here, the boundary is expected to sharpen to a discontinuity in $\mathcal{E}_{qq}$ and a corresponding divergence in the heat capacity.


\section{Conclusions and Discussion}

In summary, Markov Chain Monte Carlo simulations were performed to analyse the behavior of rotors on 2D-triangular, fcc and hcp lattices interacting through the quadrupole-quadrupole interaction.

The triangular lattice gives a structure with 3/4 molecules in plane and 1/4 perpendicular to it.  The fcc system consistently gives the P$a3$ structure upon cooling, in which each close packed plane is similar to the triangular lattice but with the atoms tilted out of plane so that all (111) planes are equivalent.

The hcp system the other hand is strongly dependent on the $c/a$ ratio. In the range of $c/a$ ratios near the ideal value of 1.633, there is significant variation in the type of stacking observed. We showed that P$ca2_1$ stacking is stable below the ideal $c/a$, whereas low-symmetry stackings of distorted P6-type layers are found at and above the ideal value. Many different types of stacking of such layers are possible with very similar energies. These P6$_3/m$ and P$ca2_1$ are among the structures predicted in hydrogen by ab-initio structure search\cite{pickard2007structure}, suggesting that quadrupole interactions are important in Phase II.  Interestingly, the Phase III candidates are not found in our MC, implying that other contributions to the free energy such as packing efficiency drive the transition to phase III\cite{magdau2017simple}.

The P$a3$ type structure is the most favorable energetically, but it is based around fcc stacking which is not favored by the van der Waals interaction\cite{loach2017stacking,jackson2002lattice}. P$a3$ is the $\alpha$ phase of nitrogen, presumably the energy gain from ordering the strong quadrupole moment overcomes the preference for hcp exhibited in $\beta$-N$_2$.

There is an EQQ-driven order-disorder phase transition in both hcp and fcc. $T_c$ is higher in fcc, reflecting the existence of a single favored energy minimum in P$a3$.  By contrast, hcp has multiple minima, permitting phase transformations between different ordered structures. Although $E_{qq}$ favors fcc, the fcc-hcp stability is primarily determined by angle-independent terms no required in this model. The single rotor, small-angle Metropolis-Hastings update works well for the present purpose, but does leave room for improvements, such as whole-layer inversions.

Recent X-ray work suggests that Phases III and IV in hydrogen are also based on an hcp lattice\cite{ji2019ultrahigh}, with non-ideal $c/a$.  Our EQQ-based model does not find the DFT candidate structures for these high-pressure phases, so we can infer that the symmetry-breakings in Phase III and IV are not the consequence of quadrupole-quadrupole interactions.

In summary, we have demonstrated that the quadrupole-quadrupole interaction alone stabilizes a number of hcp-based structures previously proposed as candidates for hydrogen Phase II, but that this interaction cannot be responsible for the preference of hcp over fcc stacking.  The actual stable structure is highly dependent on the $c/a$ ratio, with the crossover between P$ca2_1$ and P6$_3/m$-type being very close to the experimentally-observed value.
\bibliographystyle{apsrev}
\bibliography{all-revised}

\end{document}